\documentclass[preprint,12pt]{elsarticle}

 \usepackage{graphicx}
\usepackage{amssymb}
\usepackage{multirow}
\journal{Physics Letters B}

\begin{document}

\begin{frontmatter}



\title{Comparison of kaon and pion valence quark distributions in a statistical model}


\author[a,b]{Mary Alberg\corref{*}}
\cortext[*]{Corresponding author}
\ead{alberg@seattleu.edu}
\author[a]{Jeffrey Tibbals}\ead{tibbalsj@seattleu.edu}

\address[a]{Department of Physics, Seattle University, Seattle,
WA 98122, USA}

\address[b]{Department of Physics, University of Washington, Seattle,
WA 98195, USA}

\begin{abstract}
We have calculated the Bjorken-$x$ dependence of the kaon and pion valence quark distributions in a statistical
model. Each meson is described by a Fock state expansion in terms of quarks, antiquarks and gluons. Although Drell-Yan experiments have measured
the pion valence quark distributions directly, the kaon valence quark
distributions have only been deduced from the measurement of the ratio $\bar{u}_K(x)/\bar{u}_\pi(x)$. We show that, using no free parameters, our model predicts the decrease of
this ratio with increasing $x$.
\end{abstract}

\begin{keyword}

meson parton distributions \sep statistical model


\end{keyword}

\end{frontmatter}

% \linenumbers

%\section{}
%\label{}
 \section{Introduction}
A determination of parton distribution functions (PDFs) in mesons is important as a test of QCD. A Drell-Yan experiment has measured the pion valence quark distributions \cite{Conway:1989uq}, but the kaon valence quark
distributions have only been deduced from the measurement of the ratio $\bar{u}_K(x)/\bar{u}_\pi(x)$ \cite{Badier:1980rt}.
Theoretical calculations of pion PDFs have used the Dyson Schwinger equations (DSE),  the Nambu-Jona-Lasinio (NJL) model,  instantons,  constituent quark models, and statistical models.  For a recent review see Holt and Roberts  \cite{Holt:2010fk}.
Kaon PDFs have been calculated with the NJL model \cite{Shigetani:1993lr,Davidson:2001cc}, a meson cloud model \cite{Avila:2003lr}, a valon model \cite{Arash:2004fk}. and the Dyson Schwinger equations \cite{Nguyen:2011ys,Hecht:2001lr}.

In section 2 we summarize the statistical model, review our calculation for pion PDFs, and describe our calculation for kaon PDFs. In section 3 we compare our valence quark ratio  to experiment and other theoretical calculations.

 \section{statistical model}
 
Zhang and collaborators \cite{Zhang:2001kx,Zhang:2002vn,Zhang:2002uq} have used a simple statistical
model to calculate parton distribution functions in the proton. They considered the proton to be an ensemble of quark-gluon Fock states, and used detailed balance between  states  
to determine the state probabilities. A Monte-Carlo program was used to generate momentum distribution functions for each state, from which proton PDFs were determined. Their model, using no free parameters, predicted an integrated asymmetry $(\bar{d} - \bar{u})=0.124$, consistent with the experimental value $0.118 \pm 0.012$ measured in deep inelastic scattering (DIS) \cite{Amaudruz:1991qy}. Their results were also in good agreement with the $\bar{d}(x) - \bar{u}(x)$ distributions measured in DIS \cite{Ackerstaff:1998lr} and Drell-Yan experiments \cite{Baldit:1994fk, Hawker:1998lr}.

%%%%%%
\subsection{statistical model for the pion}
%%%%%%
We have used Zhang et al.'s statistical model to calculate PDFs for the pion that are in good agreement with experiment and other calculations  \cite{Alberg:2005lr}. We assumed a light sea of $\bar{u}u, \bar{d}d$ and gluons. The Fock state expansion for the pion is

\begin{equation}
|\pi^+> =\sum_{i,j,k} c_{ijk}|\{u\bar{d}\}\{ijk\}>.
\end{equation}
with $i$ the number of $\bar{u}u$ pairs, $j$ the number 
of $\bar{d}d $ pairs and $k$ the number of gluons. The leading term in the expansion, $\{ijk\}=\{000\}$,  represents the valence quark state $u\bar{d}$. The probability of finding the pion in the state
$|\{u\bar{d}\}\{ijk\}>$ is $\rho_{ijk}=|c_{ijk}|^2$, with

\begin{equation}
\sum_{i,j,k} \rho_{ijk} = 1 \; .\label{norm}
\end{equation}
Detailed balance between any two Fock states requires that 

\begin{eqnarray}
\rho_{ijk} N(|\{uud\}\{ijk\}> \rightarrow |\{uud\}\{i^\prime j^\prime k^\prime\}>) 
& \equiv & \nonumber \\
      \rho_{i^\prime j^\prime k^\prime} N(|\{uud\}\{i^\prime j^\prime k^\prime\}> 
\rightarrow |\{uud\}\{ijk\})>,&&
\end{eqnarray}
in which $N(A \rightarrow B))$ is the transfer rate of state $A$ into state $B$. Transfer rates between states are assumed to be proportional to the number of partons that can  split or recombine.  Taking into account the processes $q \leftrightarrow q\, g$, 
$g \leftrightarrow q\, \bar{q}$, and $g\leftrightarrow gg$,
\begin{equation}
\frac{\rho_{ijk}}{\rho_{000}} = \frac{1}{i!(i+1)!j!(j+1)!}\prod_{n=0}^{k-1}\frac{2+2i+2j+n}{(2+2i+2j)(n+1)+\frac{n(n+1)}{2}},
\end{equation}
This equation, together with the normalization condition (\ref{norm}), determines the $\rho_{ijk}$. The $\pi^+$ sea is flavor symmetric, since the equation is symmetric in $i$ and $j$. 

The average number of partons in the pion, $\bar{n}_\pi$, is given by
\begin{equation}
\bar{n}_\pi=\int_0^1 [u(x) + d(x)+\bar{u}(x)+\bar{d}(x) +g (x)] dx = 4.74\; .
\end{equation}

For the pion we assumed massless partons, and used the Monte Carlo event generator RAMBO \cite{Kleiss:1986fk} to determine the momentum distribution  $f_n(x)$ for each $n$-parton state, $n=2+2(i+j)+k$. The flavor distributions for each Fock state are
\begin {eqnarray}
\bar{d}_{ijk}(x) = f_n(x) (1+j)\;   , &
u_{ijk}(x) = f_n(x) (1 + i) \; ,
\end {eqnarray}
\begin {eqnarray}
\bar{u}_{ijk}(x) = f_n(x) i\;   , &
d_{ijk}(x) = f_n(x)  j  \;  ,
\end {eqnarray}
and for the gluons
\begin {equation}
g_{ijk}(x) = f_n(x)  k  \;  .
\end {equation}
The PDFs are found by summing these distribution functions over all values of $\{ijk\}$
\begin{equation}
u(x) = \sum_{i,j,k} \rho_{ijk}\, u_{ijk}(x)= \bar{d}(x) \; ,
\end{equation}
\begin{equation}
d(x) = \sum_{i,j,k} \rho_{ijk}\, d_{ijk}(x)= \bar{u}(x) \; ,
\end{equation}
and
\begin{equation}
g(x) = \sum_{i,j,k} \rho_{ijk}\, g_{ijk}(x)\; .
\end{equation}
The valence quark distribution function is
\begin{equation}
v(x) = u(x)-\bar{u}(x)=\bar{d}(x)-d(x)
\end{equation}
and the momentum distribution functions satisfy
\begin{equation}
\int_0^1 x  [u(x) + d(x)+\bar{u}(x)+\bar{d}(x) +g (x)] dx = 1 \; .
\end{equation}

\begin{figure}[htbp]
\begin{center}
\includegraphics[width=3.00in]{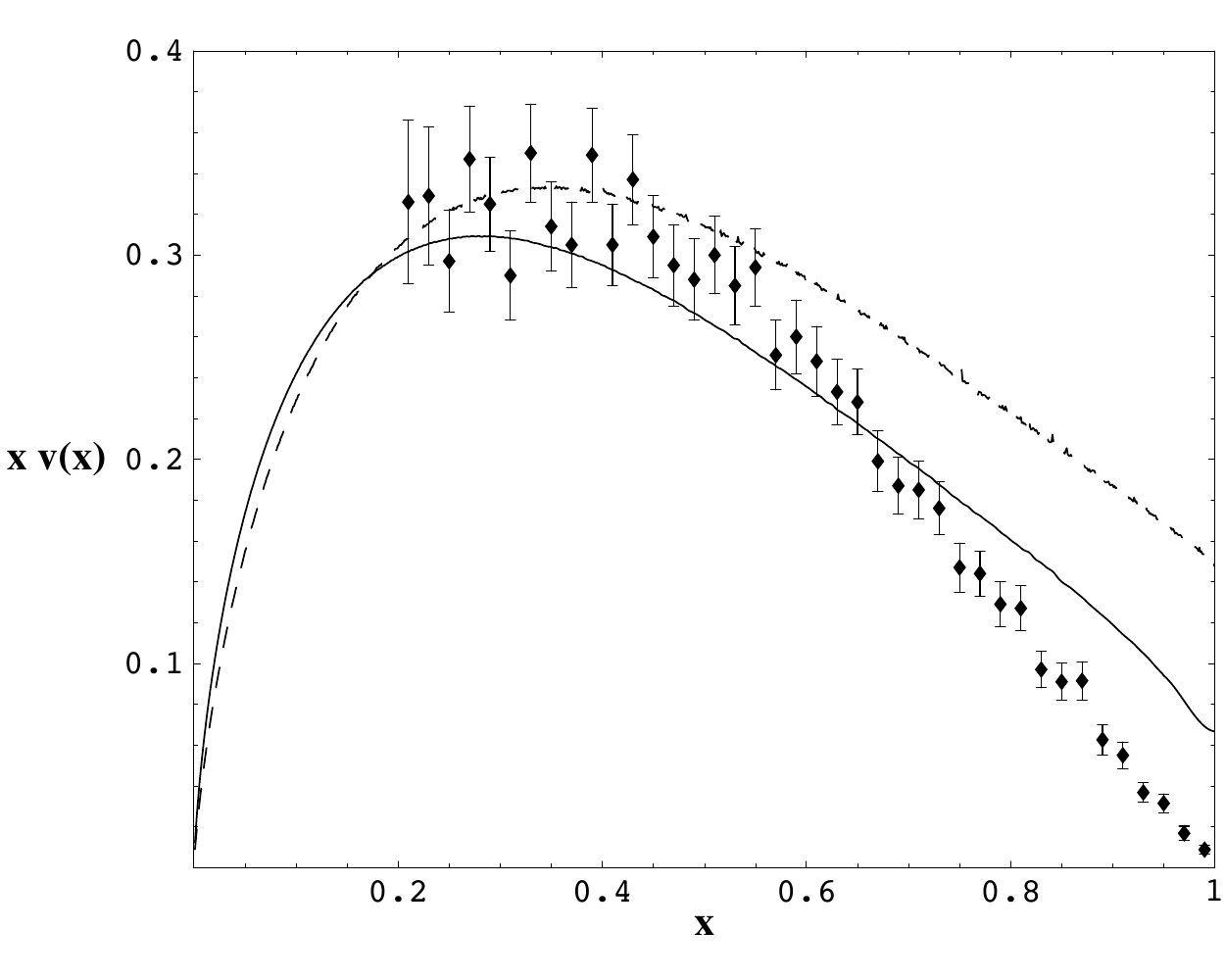}
\caption{Our results (solid curve) for the valence quark distribution $x\, v(x)$ in the pion, compared to the experimental results of Conway et al. \cite{Conway:1989uq}. The dashed curve shows our results without any evolution, which correspond to a scale of $Q_0^2 = 1.96$ GeV$^2$.
The solid curve shows our results evolved to $Q^2=16$ GeV$^2$ of the E615 experiment.}
\label{pionpdfs}
\end{center}
\end{figure}

In order to compare our valence quark distributions to experiment, we carried out an evolution in $Q^2$.  We determined the starting scale of our distributions by requiring that  the first and second moments of our valence quark distribution at $Q^2 = 4$ GeV$^2$ be equal to those found by Sutton et al. \cite{Sutton:1992qy}. This gave us a starting scale of $Q_0^2 = 1.96$ GeV$^2$. We used Miyama and Kumano's code BF1 \cite{Miyama:1996fj} for the evolution.
 Our results, shown in Fig. \ref{pionpdfs}, are in reasonable agreement with experiment. The non-zero values at $x \approx 1$ are contributed by the first term in the Fock state expansion, which consists of two massless partons,  for which the momentum distribution is constant.

 %%%%%%%%%%%%%%%%%%% INSERT PION FIGURE - compare to exp

%%%%%%
\subsection{statistical model for the kaon}
%%%%%%
As for the pion, we assume that the kaon has a light sea of $\bar{u}u, \bar{d}d$ and gluons. The Fock state expansion for the $K^+$ is
\begin{equation}
|K^+> =\sum_{i,j,k} c_{ijk}|\{u\bar{s}\}\{ijk\}>.
\end{equation}
Again, assuming that transfer rates between states are assumed to be proportional to the number of partons that can split or recombine, and including the processes $q \leftrightarrow q\, g$, 
$g \leftrightarrow q\, \bar{q}$, and $g\leftrightarrow gg$,
\begin{equation}
\frac{\rho_{ijk}}{\rho_{000}} = \frac{1}{i!(i+1)!j!^2}\prod_{n=0}^{k-1}\frac{2+2i+2j+n}{(2+2i+2j)(n+1)+\frac{n(n+1)}{2}},
\end{equation}
We find the average number of partons in the kaon $\bar{n}_K = 5.23$. The $K^+$ sea is asymmetric, with $<\bar{u}u> = 0.424$ and $<\bar{d}d> = 0.685$. As in the case of the proton, the valence $u$ quark provides more pathways for annihilation of the $\bar{u}$ quarks in the sea, whereas  the $\bar{d}$ quarks can only annihilate on other sea quarks. Detailed balance then requires an excess of $\bar{d}d$ over $\bar{u}u$.

\begin{figure}[htbp]
\begin{center}
\includegraphics[width=3.00in]{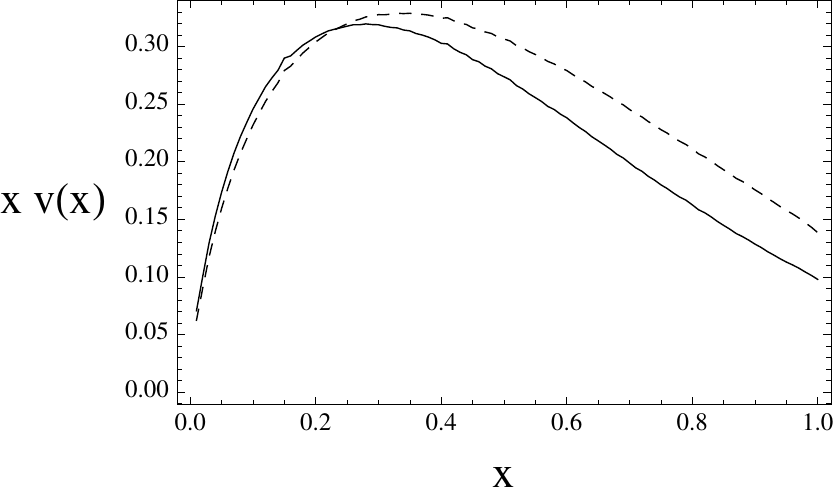}
\caption{Valence quark distributions $x\, v(x)$ for the kaon (solid curve) and the pion (dashed curve), assuming massless partons. The PDFS are calculated at our starting scale of $Q_0^2 = 1.96$ GeV$^2$.}
\label{masslesscf}
\end{center}
\end{figure}

For the kaon we also used RAMBO to determine momentum distributions for each $n$-parton state. If all partons are considered to be massless, including the $\bar{s}$, PDFS for the kaon differ from those of the pion, because   $\bar{n}_K > \bar{n}_\pi$. The kaon's momentum is shared among more partons than the pion's momentum, so the momentum fraction contributed by its valence quarks 
\begin{equation}
\int_0^1 x u_{K}(x) dx =0.46\; 
\end{equation}
is less than the momentum fraction contributed by the pion's valence quarks 
\begin{equation}
\int_0^1 x u_{\pi}(x) dx =0.50\; .
\end{equation}
This is seen in the valence PDFs shown in Fig \ref{masslesscf}. The momentum distributions at $x \approx 1$ are determined by the probability of finding the meson in the leading term of its Fock state expansion, the valence state, which is 0.14 for the pion and 0.10 for the kaon. % %%%%%%%%%%%%%%%%%%% INSERT FIGURE - comparison of massless pion and kaon PDFS 

\begin{figure}[htbp]
\begin{center}
\includegraphics[width=3.00in]{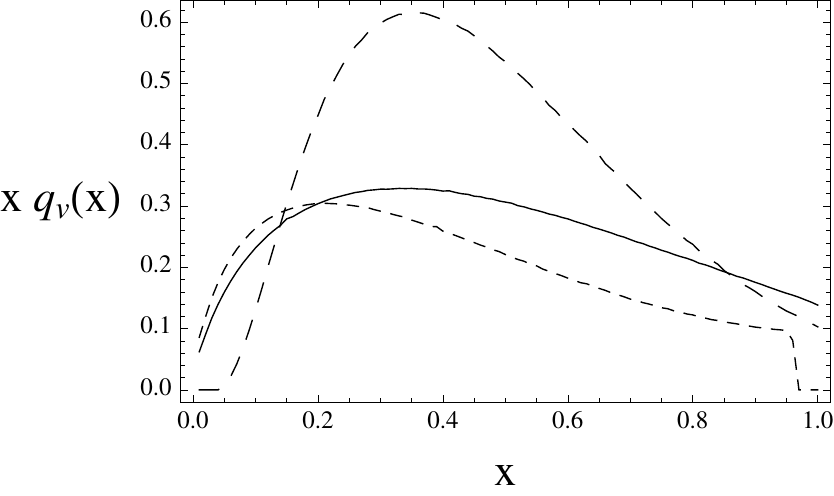}
\caption{Valence quark distributions $x\, u_{\pi}(x)$ for the pion (solid curve),  $x\,u_{K}(x)$ (short-dashed curve) and $x\,\bar{s}_{K}(x)$ (long-dashed curve) for the kaon, with  an $\bar{s}$ mass $M = 100$ MeV. The PDFS are calculated at our starting scale of $Q_0^2 = 1.96$ GeV$^2$.}
\label{massivecf}
\end{center}
\end{figure}

We also considered the case in which the $\bar{s}$ was given its current quark mass  $M = 100$ MeV, but all other partons were considered massless. We then have two distributions, $f_{nM}(x)$, for the massive $\bar{s}$, and $f_{n0}(x)$ for all the other partons, considered massless. We determined PDFs for the kaon, and compare them to the pion, in Fig. \ref{massivecf}.  As expected, the $u_{K}(x)$ distribution shifts to lower $x$, and the $\bar{s}_{K}(x)$ distribution peaks at higher $x$.
Parton numbers and momentum fractions for the pion and the kaon are shown in Table \ref{moments}.
\begin{table}[htdp]
\caption{Parton numbers and momentum fractions for the pion and the kaon, calculated at the starting scale of $Q_0^2 = 1.96$ GeV$^2$. For the kaon, $<x>_0$ is the momentum fraction calculated assuming a massless $\bar{s}$, and $<x>_M$ is the momentum fraction calculated for an $\bar{s}$ mass $M = 100$ MeV.}
\begin{center}
{\begin{tabular}{| l | lcc | lccc |}
\hline
%\multicolumn
		&		& $\pi^+$ 	  &		&		&		& K$^+$	&		\\
\hline
		& partons  & $\bar{n}$&$<x>$	& partons  & $\bar{n}$&$<x>_0$	&$<x>_M$\\
\hline
valence 	&  $u_v$ 		    &  	1	&  0.25 &    $u_v$  		& 	1	& 0.23	&0.20	\\
   		&  $\bar{d}_v$      &  	1	&  0.25 &    $\bar{s}$ 	&	 1	& 0.23	& 0.35	\\
sea     	&  $u+\bar{u}$      &  0.82  & 0.14 &     $u+\bar{u}$ 	& 	0.83	& 0.13	& 0.11	\\
	   	&  $d+\bar{d}$      &  0.82   & 0.14  &    $d+\bar{d}$ 	 & 	1.31	&  0.22	& 0.18	\\
       		&  gluons         	    &  1.10   & 0.22  &      gluons 		& 	1.09	& 0.20	&0.16	\\
total   	&  			     &  4.74  & 1.00  &    	 		& 	5.23	& 1.0		&1.00	\\

\hline
\end{tabular}}
\label{moments}
\end{center}
\end{table}
%

%%%%%%%%% INSERT FIGURE - comparison of massless pion and massive sbar for kaon PDFS 
%

%%%%%%
\section{Comparison with experimental ratio and other theoretical calculations}
\begin{figure}[htbp]
  \centerline{
    \mbox{\includegraphics[width=2.50in,height=1.60in]{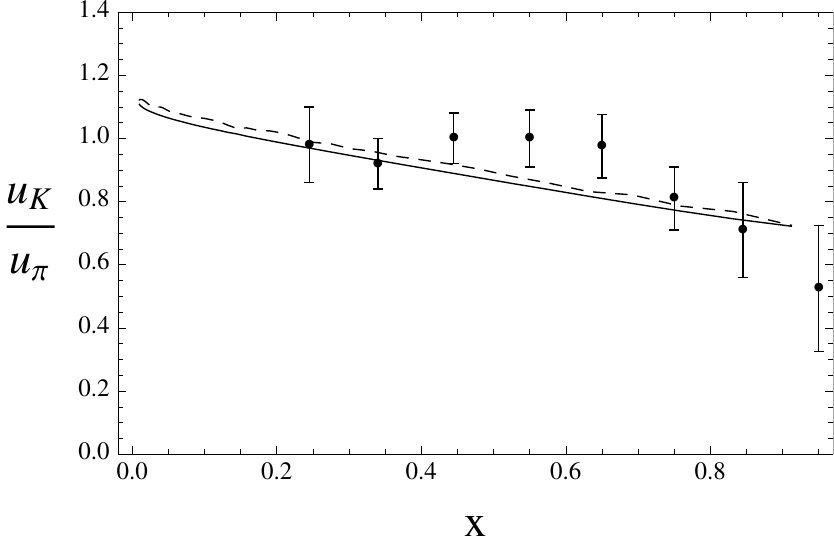}}
    \mbox{\includegraphics[width=2.50in,height=1.60in]{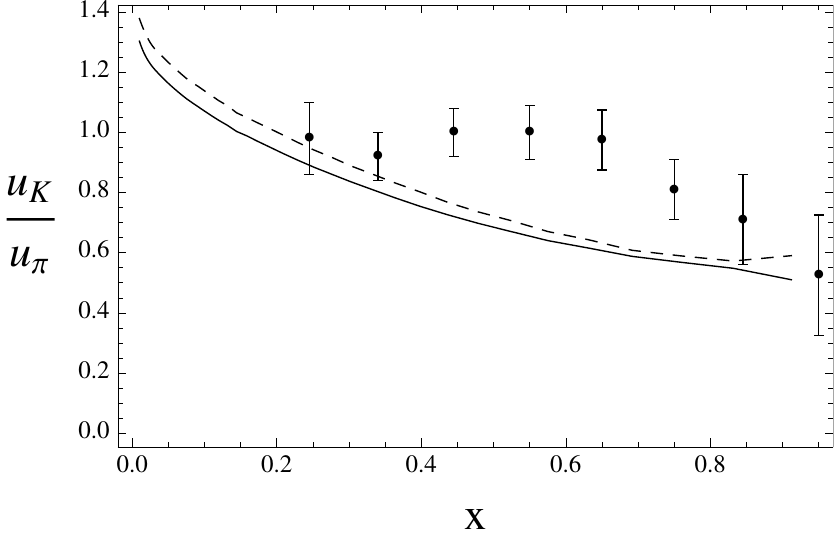}}
  }
  \caption{The valence quark distribution ratio $u_{K}(x)/u_{\pi}(x)$. The left panel shows our calculation for massless partons. In the right panel the calculation used $M = 100$ MeV for the $\bar{s}$. The dashed curves were calculated for the starting scale; the solid curves are for distributions evolved to  $Q^2=27$ GeV$^2$ for comparison with the Badier et al. \cite{Badier:1980rt} experiment. }
  \label{ratio}
  \end{figure}

\begin{figure}[htbp]
\begin{center}
\includegraphics[width=3.00in]{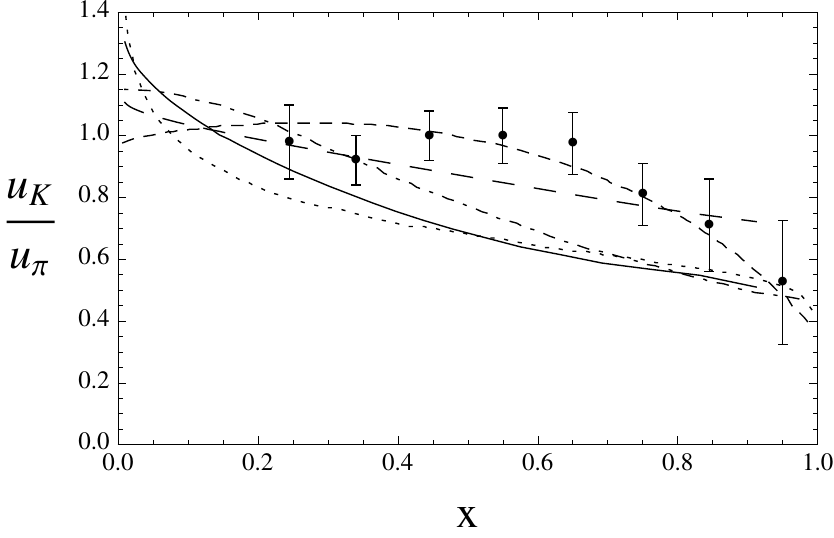}
\caption{Calculations of the valence quark distribution ratio $u_{K^+}(x)/u_{\pi^+}(x)$ compared to experiment \cite{Badier:1980rt}.  The solid line is our calculation using $M = 100$ MeV for the $\bar{s}$ mass.  The long dashed curve is our calculation using massless partons. The short-dashed line is the DSE prediction of Nguyen et al.  \cite{Nguyen:2011ys}. The dot-dash curve is the NJL calculation of Davidson and Arriola \cite{Davidson:2001cc} The dotted curve is the valon model calculation  of Arash \cite{Arash:2004fk} for a massive $\bar{s}$.  }
\label{theorycf}
\end{center}
\end{figure}

%%%%%%
Badier et al. \cite{Badier:1980rt} determined the valence quark ratio $\bar{u}_K(x)/\bar{u}_\pi(x)$ from Drell-Yan experiments with K$^-$ and $\pi^-$ beams incident on a platinum target. By symmetry this ratio is equal to the valence quark ratio $u_{K}(x)/u_{\pi}(x)$ for the positively charged mesons we have discussed above. We compare our results to experiment in Fig. \ref{ratio}. Evolution has very little effect on the ratios. 
%%%%%%%%% INSERT FIGURE - comparison of ratios for massless pion and massive sbar for kaon PDFS 
The massless parton calculation agrees better with experiment than the calculation that included the $\bar{s}$ mass. However, as seen in Table \ref{moments}, the momentum fraction carried by the $u_{K}(x)$ distribution, drops from $23\%$ (massless $\bar{s}$) to $20\%$ (massive $\bar{s}$) and the momentum distribution shifts  to lower $x$,  as seen in Fig. \ref{massivecf}. Both effects cause the ratio to increase for $x \leq 0.2$ and decrease for larger values of $x$.
In Fig. \ref{theorycf}  we compare our results to other calculations, and to experiment. 

%%%%%%%%% INSERT Comparison FIGURE 
The best agreement with experiment is the recent DSE calculation of Nguyen et al. \cite{Nguyen:2011ys}, which used a full Bethe-Salpeter amplitude. Other calculations predict an $x$-dependence  similar to ours. Davidson and Arriola, using an NJL model, found the same trend of a decreasing ratio with increasing $x$, as did the earlier NJL calculation of Shigetani et al. \cite{Shigetani:1993lr}. Arash, using a valon model with equal masses for the $u$ and $\bar{s}$ valence quarks in the kaon, found reasonable agreement with experiment, but for a massive $\bar{s}$, the ratio did not agree as well. These models do not fit the experimental values of the ratio for the mid-range of $x$ as well as the DSE calculation. The models predict a wide range of values for the ratio as $x \rightarrow 0$. Data for the ratio in the region $x < 0.2$ is needed to test the models. In their review article, Holt and Roberts \cite{Holt:2010fk} have noted that the experimental data for the ratio are `not of sufficient quality to test and verify our understanding of pion and kaon structure'. It is important to make higher statistics measurements of both the pion and kaon parton distributions, and to extend the measurements to lower $x$.
\section{Conclusions}

We have used a simple statistical model, developed for the calculation of parton distribution functions in the proton, to calculate the parton distribution functions of the pion and the kaon. We find that the ratio of  valence quark distributions, $u_{K}(x)/u_{\pi}(x)$, shows the expected decrease with increasing $x$.

\section{Acknowledgement}

We thank Michael Clement, Thomas Shelly and Bonnie Canion for computational work with RAMBO and calculation of distribution functions, and thank Ernest Henley for helpful discussions.
This work was supported in part by the U.S. Department of Energy and the Research in Undergraduate Institutions program of the U.S. National Science Foundation, Grant No.0855656. 


\bibliographystyle{model1-num-names}
\bibliography{K_pi}

\begin{thebibliography}{20}
\expandafter\ifx\csname natexlab\endcsname\relax\def\natexlab#1{#1}\fi
\providecommand{\bibinfo}[2]{#2}
\ifx\xfnm\relax \def\xfnm[#1]{\unskip,\space#1}\fi
%Type = Article
\bibitem[{Conway et~al.(1989)}]{Conway:1989uq}
\bibinfo{author}{J.~S. Conway}, et~al.,
\newblock \bibinfo{title}{Experimental study of muon pairs produced by 252-gev
  pions on tungsten},
\newblock \bibinfo{journal}{Phys. Rev. D} \bibinfo{volume}{39}
  (\bibinfo{year}{1989}) \bibinfo{pages}{92--122}.
%Type = Article
\bibitem[{Badier et~al.(1980)}]{Badier:1980rt}
\bibinfo{author}{J.~Badier}, et~al.,
\newblock \bibinfo{title}{Measurement of the k-/[pi]- structure function ratio
  using the drell-yan process},
\newblock \bibinfo{journal}{Physics Letters B} \bibinfo{volume}{93}
  (\bibinfo{year}{1980}) \bibinfo{pages}{354 -- 356}.
%Type = Article
\bibitem[{Holt and Roberts(2010)}]{Holt:2010fk}
\bibinfo{author}{R.~J. Holt}, \bibinfo{author}{C.~D. Roberts},
\newblock \bibinfo{title}{Nucleon and pion distribution functions in the
  valence region},
\newblock \bibinfo{journal}{Rev. Mod. Phys.} \bibinfo{volume}{82}
  (\bibinfo{year}{2010}) \bibinfo{pages}{2991--3044}.
%Type = Article
\bibitem[{Shigetani et~al.(1993)Shigetani, Suzuki, and Toki}]{Shigetani:1993lr}
\bibinfo{author}{T.~Shigetani}, \bibinfo{author}{K.~Suzuki},
  \bibinfo{author}{H.~Toki},
\newblock \bibinfo{title}{Pion structure function in the nambu and jona-lasinio
  model},
\newblock \bibinfo{journal}{Physics Letters B} \bibinfo{volume}{308}
  (\bibinfo{year}{1993}) \bibinfo{pages}{383 -- 388}.
%Type = Article
\bibitem[{Davidson and Ruiz~Arriola(2002)}]{Davidson:2001cc}
\bibinfo{author}{R.~M. Davidson}, \bibinfo{author}{E.~Ruiz~Arriola},
\newblock \bibinfo{title}{{Parton distributions functions of pion, kaon and eta
  pseudoscalar mesons in the NJL model}},
\newblock \bibinfo{journal}{Acta Phys. Polon.} \bibinfo{volume}{B33}
  (\bibinfo{year}{2002}) \bibinfo{pages}{1791--1808}.
%Type = Article
\bibitem[{Avila et~al.(2003)Avila, Sanabria, and Magnin}]{Avila:2003lr}
\bibinfo{author}{C.~Avila}, \bibinfo{author}{J.~C. Sanabria},
  \bibinfo{author}{J.~Magnin},
\newblock \bibinfo{title}{Pion and kaon parton distribution functions in a
  meson-cloud model},
\newblock \bibinfo{journal}{Phys. Rev. D} \bibinfo{volume}{67}
  (\bibinfo{year}{2003}) \bibinfo{pages}{034022}.
%Type = Article
\bibitem[{Arash(2004)}]{Arash:2004fk}
\bibinfo{author}{F.~Arash},
\newblock \bibinfo{title}{Meson structure functions in the valon model},
\newblock \bibinfo{journal}{Phys. Rev. D} \bibinfo{volume}{69}
  (\bibinfo{year}{2004}) \bibinfo{pages}{054024}.
%Type = Article
\bibitem[{Nguyen et~al.(2011)Nguyen, Bashir, Roberts, and
  Tandy}]{Nguyen:2011ys}
\bibinfo{author}{T.~Nguyen}, \bibinfo{author}{A.~Bashir},
  \bibinfo{author}{C.~D. Roberts}, \bibinfo{author}{P.~C. Tandy},
\newblock \bibinfo{title}{Pion and kaon valence-quark parton distribution
  functions},
\newblock \bibinfo{journal}{Phys. Rev. C} \bibinfo{volume}{83}
  (\bibinfo{year}{2011}) \bibinfo{pages}{062201}.
%Type = Article
\bibitem[{Hecht et~al.(2001)Hecht, Roberts, and Schmidt}]{Hecht:2001lr}
\bibinfo{author}{M.~B. Hecht}, \bibinfo{author}{C.~D. Roberts},
  \bibinfo{author}{S.~M. Schmidt},
\newblock \bibinfo{title}{Valence-quark distributions in the pion},
\newblock \bibinfo{journal}{Phys. Rev. C} \bibinfo{volume}{63}
  (\bibinfo{year}{2001}) \bibinfo{pages}{025213}.
%Type = Article
\bibitem[{Zhang et~al.(2001)Zhang, Zhang, and Ma}]{Zhang:2001kx}
\bibinfo{author}{Y.-J. Zhang}, \bibinfo{author}{B.~Zhang},
  \bibinfo{author}{B.-Q. Ma},
\newblock \bibinfo{title}{Detailed balance and sea-quark flavor asymmetry of
  proton},
\newblock \bibinfo{journal}{Physics Letters B} \bibinfo{volume}{523}
  (\bibinfo{year}{2001}) \bibinfo{pages}{260 -- 264}.
%Type = Article
\bibitem[{Zhang et~al.(2002{\natexlab{a}})Zhang, Zou, and Yang}]{Zhang:2002vn}
\bibinfo{author}{Y.-J. Zhang}, \bibinfo{author}{B.-S. Zou},
  \bibinfo{author}{L.-M. Yang},
\newblock \bibinfo{title}{Parton distribution of proton in a simple statistical
  model},
\newblock \bibinfo{journal}{Physics Letters B} \bibinfo{volume}{528}
  (\bibinfo{year}{2002}{\natexlab{a}}) \bibinfo{pages}{228 -- 232}.
%Type = Article
\bibitem[{Zhang et~al.(2002{\natexlab{b}})Zhang, Deng, and Ma}]{Zhang:2002uq}
\bibinfo{author}{Y.-J. Zhang}, \bibinfo{author}{W.-Z. Deng},
  \bibinfo{author}{B.-Q. Ma},
\newblock \bibinfo{title}{Principle of balance and the sea content of the
  proton},
\newblock \bibinfo{journal}{Phys. Rev. D} \bibinfo{volume}{65}
  (\bibinfo{year}{2002}{\natexlab{b}}) \bibinfo{pages}{114005}.
%Type = Article
\bibitem[{Amaudruz et~al.(1991)}]{Amaudruz:1991qy}
\bibinfo{author}{P.~Amaudruz}, et~al.,
\newblock \bibinfo{title}{Gottfried sum from the ratio
  ${\mathit{f}}_{2}^{\mathit{n}}$/${\mathit{f}}_{2}^{\mathit{p}}$},
\newblock \bibinfo{journal}{Phys. Rev. Lett.} \bibinfo{volume}{66}
  (\bibinfo{year}{1991}) \bibinfo{pages}{2712--2715}.
%Type = Article
\bibitem[{Ackerstaff et~al.(1998)}]{Ackerstaff:1998lr}
\bibinfo{author}{K.~Ackerstaff}, et~al.,
\newblock \bibinfo{title}{Flavor asymmetry of the light quark sea from
  semi-inclusive deep-inelastic scattering},
\newblock \bibinfo{journal}{Phys. Rev. Lett.} \bibinfo{volume}{81}
  (\bibinfo{year}{1998}) \bibinfo{pages}{5519--5523}.
%Type = Article
\bibitem[{Baldit et~al.(1994)}]{Baldit:1994fk}
\bibinfo{author}{A.~Baldit}, et~al.,
\newblock \bibinfo{title}{Study of the isospin symmetry breaking in the light
  quark sea of the nucleon from the drell-yan process},
\newblock \bibinfo{journal}{Physics Letters B} \bibinfo{volume}{332}
  (\bibinfo{year}{1994}) \bibinfo{pages}{244 -- 250}.
%Type = Article
\bibitem[{Hawker et~al.(1998)}]{Hawker:1998lr}
\bibinfo{author}{E.~A. Hawker}, et~al.,
\newblock \bibinfo{title}{Measurement of the light antiquark flavor asymmetry
  in the nucleon sea},
\newblock \bibinfo{journal}{Phys. Rev. Lett.} \bibinfo{volume}{80}
  (\bibinfo{year}{1998}) \bibinfo{pages}{3715--3718}.
%Type = Article
\bibitem[{Alberg and Henley(2005)}]{Alberg:2005lr}
\bibinfo{author}{M.~Alberg}, \bibinfo{author}{E.~M. Henley},
\newblock \bibinfo{title}{Parton distributions in the proton and pion},
\newblock \bibinfo{journal}{Physics Letters B} \bibinfo{volume}{611}
  (\bibinfo{year}{2005}) \bibinfo{pages}{111 -- 115}.
%Type = Article
\bibitem[{Kleiss et~al.(1986)Kleiss, Stirling, and Ellis}]{Kleiss:1986fk}
\bibinfo{author}{R.~Kleiss}, \bibinfo{author}{W.~J. Stirling},
  \bibinfo{author}{S.~D. Ellis},
\newblock \bibinfo{title}{A new monte carlo treatment of multiparticle phase
  space at high energies},
\newblock \bibinfo{journal}{Computer Physics Communications}
  \bibinfo{volume}{40} (\bibinfo{year}{1986}) \bibinfo{pages}{359 -- 373}.
%Type = Article
\bibitem[{Sutton et~al.(1992)Sutton, Martin, Roberts, and
  Stirling}]{Sutton:1992qy}
\bibinfo{author}{P.~J. Sutton}, \bibinfo{author}{A.~D. Martin},
  \bibinfo{author}{R.~G. Roberts}, \bibinfo{author}{W.~J. Stirling},
\newblock \bibinfo{title}{Parton distributions for the pion extracted from
  drell-yan and prompt photon experiments},
\newblock \bibinfo{journal}{Phys. Rev. D} \bibinfo{volume}{45}
  (\bibinfo{year}{1992}) \bibinfo{pages}{2349--2359}.
%Type = Article
\bibitem[{Miyama and Kumano(1996)}]{Miyama:1996fj}
\bibinfo{author}{M.~Miyama}, \bibinfo{author}{S.~Kumano},
\newblock \bibinfo{title}{Numerical solution of q2 evolution equations in a
  brute-force method},
\newblock \bibinfo{journal}{Computer Physics Communications}
  \bibinfo{volume}{94} (\bibinfo{year}{1996}) \bibinfo{pages}{185 -- 215}.

\end{thebibliography}
\end{document}